\documentstyle[mprocl]{article}
\def\Journal#1#2#3#4{{#1} {\bf #2}, #3 (#4)}
\def\AP{{\em Ann. Phys.} (N.Y.)}
\def\PR{\em Phys. Rep.}
\def\PLB{{\em Phys. Lett.} B}
\def\JMP{\em J. Math. Phys.}

\begin{document}

\title{TOWARDS HYPERGRAVITY}

\author{Djordje \v Sija\v cki}
\address{Institute of Physics, P.O. Box 57, \\
11001 Belgrade, Yugoslavia}

\maketitle\abstracts{
The aim of this paper is to discuss a kinematical algebraic structure of a
theory of gravity, that would be unitary, renormalizable and coupled in the
same manner to both spinorial and tensorial matter fields. An analysis of
the common features as well as differences of the Yang-Mills theories and
gauge theories of gravity is carried out. In particular, we consider the
following issues: (i) Representations of the relevant global symmetry on
states and on fields, (ii) Relations between the relevant global and local
symmetries, (iii) Representations of the local symmetries on states and
fields, (iv) Dimensional analysis of the gauge algebra generators and the
number of counter terms, and (v) Coupling to the spinorial matter fields. We
conclude, that various difficulties on the gravity side can be overcome by
considering the below outlined Hypergravity gauge theory, that to some
extend parallels the string/membrane theories. This theory is based on an
infinite Lie algebraic structure of generators that transforms as "states"
of the infinite-dimensional irreducible representation of the
$\overline{SL}(4,R)$ subgroup of the Group of General Coordinate
Transformations of $R^4$. The metric-affine and Poincar\'e gauge
theories of gravity are obtained through a spontaneous symmetry
breaking mechanism, with the metric field as nonlinear symmetry realizer.}

The task of formulating a unitary renormalizable theory of Quantum
Gravity is apparently the most outstanding one in contemporary
Theoretical Physics. Numerous various field-theoretic approaches have
been proposed, however each one fails to meet all basic physical and/or
mathematical requirements. Simultaneous feature of unitary and
renormalizability of such theories has been one of the most difficult
requirements. We discuss, in the following, certain number of relevant
algebraic questions in the Yang-Mills and Gauge Theories of Gravity,
and suggest a Theory of Gravity that could evade common algebraic
obstacles of the known Theories of Gravity.

\section{Global Symmetry Representations on States and on Fields}

There is a significant difference between the representations of the
internal and space-time symmetries already at the level of global
symmetries. Representations of an internal symmetry on states and on
fields are mutually equivalent. They are of the same dimensionality,
and both are unitary representations. In the case of global space-time
symmetries, the representations on states are essentially (besides
straightforward unitary infinite-dimensional 3-translation
representations) given by the relevant little group unitary
representations. In the Poincar\'e case the relevant little groups are
either $\overline{SO}(1,3)$ or $\overline{E}(2) \supset U(1)$, yielding
$2J+1$ or 1-dimensional unitary representations respectively. In the
case of Affine symmetry, the relevant little groups are
$\overline{SL}(3,R)$ or $\tilde T_3\wedge\overline{SL}(2,R)$. The
corresponding unitary representations are infinite-dimensional. The
representations on fields are given by the representations of the non
Abelian subgroup, i.e. by $\overline{SO}(1,3)$ in the Poincar\'e case,
and $\overline{SL}(4,R)$ in the Affine case. The former ones are finite
dimensional and non-unitary, while the latter ones are
infinite-dimensional and unitary.

\section{Global vs. Local Symmetries}

The gauged version of a global internal symmetry group is given by a
continuous direct product of groups that have identical Lie algebra
structure as the starting global symmetry. In other words $[X_a, X_b] =
if_{ab}{}^c X_c$ $\to$ $[X_a(x), X_b(x)] = if_{ab}{}^c X_c(x)$, $a,b,c =
1,2,\cdots,n$, with the same structure constants. The gauging of a
space-time symmetry yields a local space-time symmetry that, as a rule,
differs considerably from the starting global symmetry. Let us consider
the Affine symmetry generated by the momentum generators $P_a$ and the
shear generators $Q_{ab}$, $a,b = 0,1,2,3$. The Poincar\'e symmetry
subgroup is generated by $P_a$ and $M_{ab} = Q_{[ab]}$. In the process
of gauging one introduces the parallel transport operators $D_a$, that
replace the translations generators $P_a$. The commutation relations
for $[Q_{ab}(x),Q_{cd}(x)]$ and for $[Q_{ab}(x),D_c(x)]$ are, up to $x$
dependence, isomorphic to the corresponding global ones,
$[Q_{ab},Q_{cd}]$ and $[Q_{ab}, P_c]$ respectively. However, the
commutator $[P_a,P_b] = 0$ is replaced in the local case by$^{1\ 2}$
$[D_a(x),D_b(x)] = e_a{}^\mu e_b{}^\nu (R^{cd}{}_{\mu\nu}(x)Q_{cd} 
(x)-T^c{}_{\mu\nu}(x)D_c(x))$. Here, not only that the type of the
commutation relations is changed, but instead of the structure
constants one has structure functions - the affine curvature $R^{ab}$
and torsion $T^a$. Thus, by gauging a finite-parameter space-time Lie
group one obtains an infinite-parameter Lie group, as most easily seen
by expanding $R^{ab}$ and $T^a$ in power series.

\section{Local Symmetry Representations on States and on Fields}

Owing to the fact that global and local internal symmetries are (up to
$x$ dependence) isomorphic, the corresponding representations on states
and on fields are all mutually equivalent. The inhomogeneous
transformation law in the local case relates to the space-time
dependence only. The local space-time symmetry representations on
states is achieved, by first selecting a form of the asymptotic
space-time (e.g. Minkowskian), thus defining a certain subgroup (e.g.
Poincar\'e), and then realizing non-linearly the infinite-parameter
local symmetry group over the states of the starting global symmetry.
As for the representations on fields, one makes use of the non-linear
representations of the local symmetry group in the space of
finite-dimensional world tensors. Representations on states and on
fields are again a prior unrelated, and there are quite a number of
unphysical field components due to the manifest covariance requirement.

\section{Local Symmetry Generators Dimension}

In contradistinction to the internal symmetry generators that are
dimensionless, the generators of the local space-time symmetry group
carry nontrivial dimension as easily seen from the Ogievetsky algebra
generators expressions $\{ P_\mu, F^{\nu_{1},\nu_{2} \cdots \nu_{n}}_\mu 
\sim x^{\nu_{1}}x^{\nu_{2}} \cdots x^{\nu_{n}} P_\mu \}$, i.e. 
$[P_\mu] = l^{-1}, [F^{\nu_{1},\nu_{2}\cdots\nu_{n}}_\mu] = l^{n-1}$. The 
$F^\nu_\mu = e^a_\mu e^{b\nu}Q_{ab}$ are dimensionless only. The non-trivial
dimensionality of the local space-time symmetry generators is related
to the appearance of dimensional coupling constants as well as of
various types of counter terms in the quantum case. Moreover, the
non-trivial field-dependent integration measure, due to non-linear
realization of the full infinite symmetry, yields an infinite number of
possible vertices, and thus an a prior non-renormalizable quantum theory.

\section{Tensorial and Spinorial Matter Fields}

In contradistinction to the case of internal symmetries, where all
representations are a priori on the same footing, there has been quite a
confusion even about the very existence of the spinorial representations of
the local space-time symmetries$^{3\ 4}$. The "world spinorial"
representations$^{5\ 6}$ do exist and have been explicitly constructed in
terms of the $\overline{SL}(4,R)$ subgroup infinite-component spinorial
representations.

\section{Hypergravity}

Consider, for simplicity, the $\overline{SL}(3,R) \subset
\overline{Diff}(3,R)$ group generated by the angular momentum $J$ and shear
$T$. The states of the corresponding infinite-dimensional
representations$^{7}$ are labeled by $(j;km)$, $j = 1,2, \cdots , |k,m| 
\le j$. We introduce an infinite set of operators $L_A = L_{(j \ k \ m)}$, and
require a Lie algebra closure among them. The resulting algebra is given by:

$[L_{(j^{\prime}\ k^{\prime}\ m^{\prime})},\  
L_{(j^{\prime\prime}\ k^{\prime\prime}\ m^{\prime\prime})}] = $
$$ 
\sum_{j \ k \ m}\left( 1-(-1)^{j^{\prime}+j^{\prime\prime}-j} \right)
\Delta(j\ j^\prime\ j^{\prime\prime})
\left( \begin{array}{ccc} j & j^\prime & j^{\prime\prime} \\
-k & k^\prime & k^{\prime\prime}\end{array}\right)
\left( \begin{array}{ccc} j & j^\prime & j^{\prime\prime} \\
-m & m^\prime & m^{\prime\prime}\end{array}\right)
L_{(j \ k \ m)}
$$

To each generator we assign a gauge potential $\Gamma^A_\mu$ and define
the corresponding Lie-algebra valued covariant derivative, as well as
the corresponding field strengths $R^A_{\mu\nu}$. In this manner we
obtain a Yang-Mills-like theory of gravity based on an infinite Lie
algebra that realizes linearly the group of General Coordinate
Transformations. As for the unitarity question we point out that the
representations on states and on fields in this theory are directly
related, while the renormalizability is boiled down to a finite number
of counter terms - the infinite Lie algebra of the theory interrelates
all possible vertices.

\end{document}